# USDA Forecasts: A meta-analysis study

**Bahram Sanginabadi**

December 2017
Economics Department
University of Hawaii at Manoa
bahram@hawaii.edu

## Abstract

The primary goal of this study is doing a meta-analysis research on two groups of published studies. First, the ones that focus on the evaluation of the United States Department of Agriculture (USDA) forecasts and second, the ones that evaluate the market reactions to the USDA forecasts. We investigate four questions. 1) How the studies evaluate the accuracy of the USDA forecasts? 2) How they evaluate the market reactions to the USDA forecasts? 3) Is there any heterogeneity in the results of the mentioned studies? 4) Is there any publication bias? About the first question, while some researchers argue that the forecasts are unbiased, most of them maintain that they are biased, inefficient, not optimal, or not rational. About the second question, while a few studies claim that the forecasts are not newsworthy, most of them maintain that they are newsworthy, provide useful information, and cause market reactions. About the third and the fourth questions, based on our findings, there are some clues that the results of the studies are heterogeneous, but we couldn't find enough evidences of publication bias.

## Introduction

Meta-analysis is a systematic approach to analyze literature review by statistical methods where the goal is to compile and contrast the findings of several related studies. For the first time, this method proposed by Glass (1976). Also, Stanley & Jarrell (1989), Walsh et al. (1989, 1990), Jarrell & Stanley (1990) are among the first researchers who applied meta-analysis. The studies that aim to aggregate and synthesize the literature on a certain topic progressively apply meta-analysis (Olkin,1995). Currently researchers apply this method in many different areas including psychology, education, science, marketing, and social sciences. Meta-analysis is quite popular among economists as well.

In this paper we do Meta-analysis while we exclusively focus on two types of studies as our inputs. First, the studies that evaluate the United States Department of Agriculture (USDA) forecasts. Second, the ones that evaluate the market reactions to these forecasts. It's important to mention that almost all the studies that focus on the USDA forecasts can be categorized in one or both of mentioned categories above. We believe it's important to get this research done because the number of published papers in these areas are quite high and they report a variety of findings which in many cases contradict each other.

We are interested in finding answers for four questions. First, how the academic published studies evaluate accuracy of USDA forecasts? In other words, do their findings show that USDA forecasts are accurate? Second, how the academic published studies evaluate market reactions to the USDA forecasts? Third, are results of the academic papers heterogeneous? Fourth, are there any clues publication bias?

In the rest of this paper, we focus on answering the mentioned questions above. In the next section, we briefly talk about the USDA forecasts. 'Methodology of data-analyzing' is the next thing that we discuss. Then, we represent 'Analysis', 'Accuracy of the USDA Forecasts', 'Market Reactions to the USDA Forecasts', 'Meta-analysis', and 'Discussion' respectively.

## The USDA Forecasts

USDA provides the monthly report "World Agricultural Supply and Demand Estimates" (WASDE) which is a comprehensive forecast of supply and demand for major crops (produced in U.S. and the rest of the world) and livestock (U.S. only). WASDE report applies the statistical reports compiled by the USDA agencies and other government agencies (Xiao et al., 2014).

## Literature Search and Data Collection

In a comprehensive search in the literature we found 54 relevant studies. We mainly applied the key words "USDA forecast", "USDA", "forecast", "Evaluation", "Accuracy", "market reaction", "market participants", etc. Search for the studies is done from November 15th to December 8th, 2015. The searching process has been done mainly through UH Manoa Library[1], Google Scholar[2], and ScienceDirect[3] websites. Fig. 1. represents the scatter plot that shows the number of published papers each year.

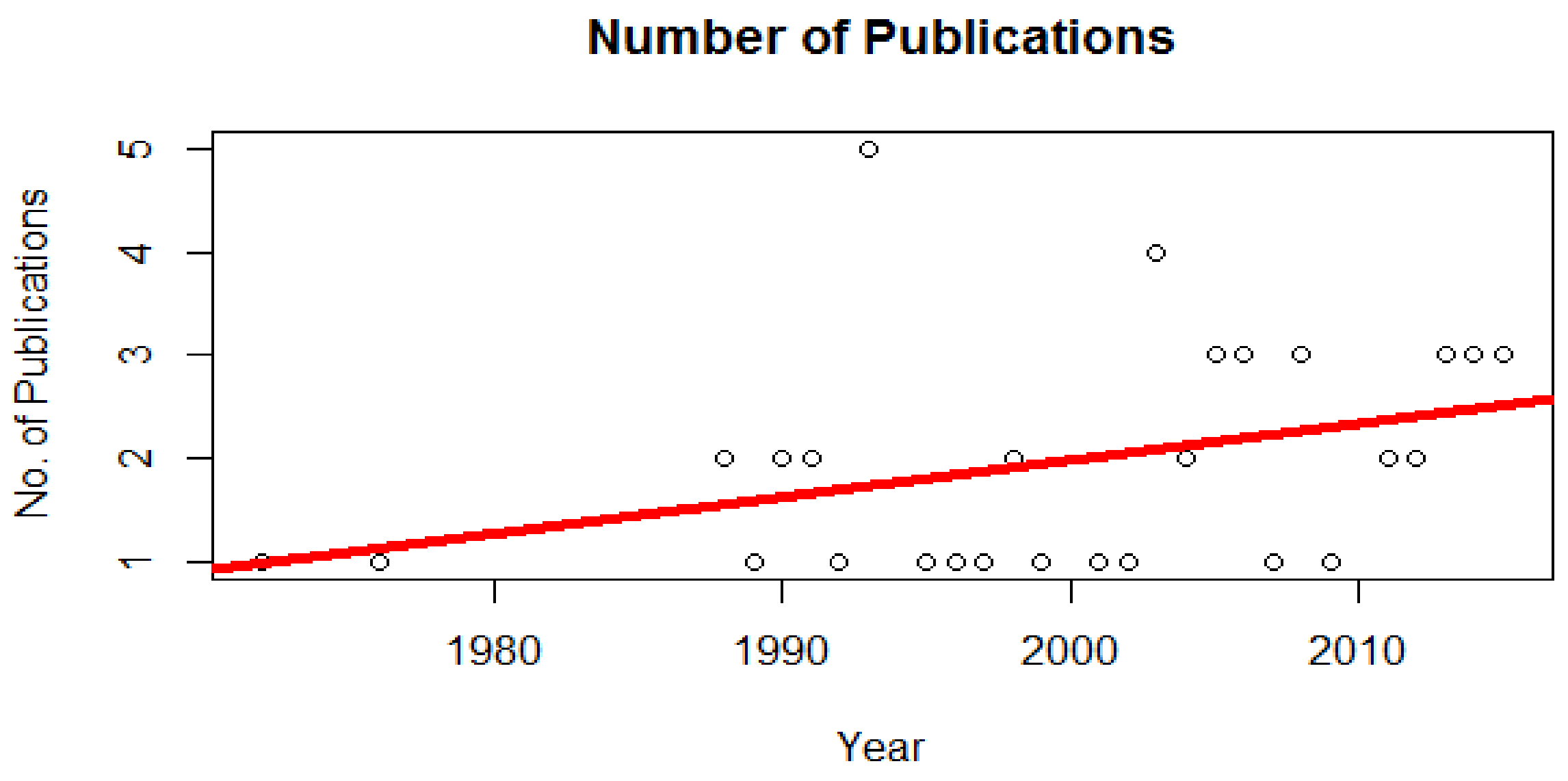

**Fig. 1. Scatter plot of Number of relevant publications**. Each dot shows the number of publications in one specific year. Note, the positive slope of the red line shows that the number of publication per year is increasing.

---

[1] http://library.manoa.hawaii.edu/
[2] https://scholar-google-com.eres.library.manoa.hawaii.edu/
[3] http://www.sciencedirect.com.eres.library.manoa.hawaii.edu/

## Methodology of Data-analyzing

To answer the first and the second questions, we summarize the findings of the relevant studies, and then we refine the results to find the patterns of their findings. To do meta-analysis we apply the metaphor package which provides functions to do the analysis in R. The package enables us to study the fixed and random effect models (Viechtbauer, 2010). Then we test for heterogeneity and publication bias which enable us to tackle the third and the fourth questions.

## Analysis

In this section, first, we provide the summary of findings of the studies that evaluate the USDA forecasts, then we summarize the findings of the ones that evaluate market reactions to the USDA forecasts. Then, in the nest section, we put all the major findings in a nutshell. Eventually, we represent meta-analysis.

| **Researcher & Topic (Accuracy of the USDA Forecasts)** | **Summary of study** |
|---|---|
| Egelkraut et al. (2003). An evaluation of crop forecast accuracy for corn and soybeans: USDA and private information agencies. | Even though, all agencies' forecast accuracy is improved and relative accuracy is varied by crop and time, the USDA predictions are more accurate than other agencies. However, when it comes to soybeans the forecast errors are very similar for all agencies. |
| Good & Irwin (2005). Understanding USDA corn and soybean production forecasts: Methods, performance and market impacts over 1970-2005. | The USDA production forecast errors are largest in August. For August, the private market forecasts for soybeans are more accurate than the USDA forecasts, but the USDA corn production forecasts are more accurate than the private market. In addition, as the growing season goes on the accuracy of the USDA forecast for soybeans improves. |
| Gunnelson et al. (1972). Analysis of the accuracy of USDA crop forecasts. | The USDA forecasts are improved moderately over 1929 to 1970, but it still underestimates the crop size, year to year production changes, and its own errors in earlier forecasts when it revises the new forecasts. |
| Irwin et al. (2014). Evaluation of Selected USDA WAOB and NASS Forecasts and Estimates in Corn and Soybeans. | Neither for corn nor for soybeans the accuracy of the WAOB forecasts have not changed significantly over time. Also, there is no evidence of bias in NASS forecasts for corn. In addition, there is some evidences of improvement in the accuracy of NASS corn forecasts over time. However, soybean forecasts usually underestimate the yield. |
| Isengildina-Massa et al. (2013). Do Big Crops Get Bigger and Small Crops Get Smaller? Further Evidence on Smoothing in US Department of Agriculture Forecasts. | The USDA forecasts for both soybeans and corn increase in big crop years and decrease in small crop years and the magnitude of smoothing is significantly large. |
| Isengildina-Massa et al. (2006). Are Revisions to USDA Crop Production Forecasts Smoothed? | The USDA forecasts are smoothed, but due to smoothing, loss in forecast accuracy happens which is statistically and economically significant in several cases. |
| Isengildina-Massa et al. (2011). Empirical confidence intervals for USDA commodity price forecasts. | This study suggests that empirical approaches such as kernel density, quantile distribution, and best fitting parametric distribution methods might be used to construct more accurate confidence intervals for USDA wheat, soybean, and corn forecasts. |
| Isengildina-Massa et al. (2013). When do the USDA forecasters make mistakes? | The errors in ending stocks forecasts are usually driven by errors in production forecasts across all commodities. In addition, for all commodities, errors in price forecasts are caused by errors in U.S. ending stocks forecasts. |
| Isengildina-Massa et al. (2012). A comprehensive evaluation of USDA cotton forecasts. | The USDA forecast overestimates China's exports, but underestimates China's domestic use and rest of the world imports. In addition, USDA repeats errors in ROW (i.e. rest of the world except China) production forecasts and overcorrects errors in ROW exports forecasts. |
| Isengildina-Massa et al. (2011). What Can We Learn from our Mistakes? Evaluating the Benefits of Correcting Inefficiencies in USDA Cotton Forecasts. | Correction for correlation in forecast revisions does not improve the USDA cotton forecasts. Correction for correlation of errors with previous year's errors and correlation of errors with forecast levels, result in improvement of USDA cotton forecasts. |
| Kastens et al. (1998). Evaluation of extension and USDA price and production forecasts. | For livestock series, Extension forecasts are more accurate than the USDA forecasts, but for the crops USDA forecasts are more accurate. However, in most of the cases Composite forecasts are more accurate than both of Extension and the USDA forecasts. |
| Manfredo & Sanders (2004). The value of public price forecasts: Additional evidence in the live Hogs market. | The lean Hogs futures-based forecast is more accurate than Extension and the USDA forecasts. |
| Meyer & Lawrence (1988). Comparing USDA Hogs and Pigs Reports to Subsequent Slaughter: Does Systematic Error Exist? | Seasonal nature of Hogs production must be scrutinized. Pigs and Hogs forecasts over emphasize this seasonality. |
| No & Salassi (2009). A sequential rationality test of USDA preliminary price estimates for selected program crops: Rice, soybeans, and wheat. | Even though, the USDA estimates are unbiased in the short-run, but they are not rational in the long-run. |
| Sanders & Manfredo (2002). USDA production forecasts for pork, beef, and broilers: an evaluation. | The USDA forecasts are unbiased, but they are not efficient. The reason is USDA do not completely consider the information from the previous forecasts. |
| Sanders & Manfredo (2003a). USDA livestock price forecasts: A | The USDA forecasts are not optimal. Broiler price forecast is biased and overall all the forecasts repeat errors. |
| comprehensive evaluation. | |
| Sanders & Manfredo (2005). A Test of Forecast Consistency Using USDA Livestock Price Forecasts. | The USDA quarterly livestock price forecasts are not consistent in the long-run. |
| Sanders & Manfredo (2008). Multiple horizons and information in USDA production forecasts. | Although the USDA forecasts are not rational they provide useful information for their users. Likewise, turkey and milk forecasts show the most consistent performance, but beef provides little information. |
| Sanders & Manfredo (2003b). Keep up the good work? An evaluation of the USDA's livestock price forecasts. | USDA Broiler price forecasts are biased. Overall, the USDA price forecasts are not optimal, and almost in all the forecasts it repeats errors. |
| Schaefer & Myers (1999). Forecasting accuracy, rational expectations, and market efficiency in the US beef cattle industry. | The USDA forecasts are inefficient and biased. |
| Von Bailey & Brorsen (1998). Trends in the accuracy of USDA production forecasts for beef and pork. | The USDA forecast underestimates production in the 1980s, but the bias disappears later. So, the accuracy of the forecasts is improved and even though the USDA forecasts are not optimal in 1980s, they show optimality after then. |
| Xiao et al. (2014). USDA and private analysts' forecasts of ending stocks: how good are they? | The USDA forecasts are unbiased, but both of the USDA and private forecasts are inefficient. Also, the accuracy of both of the USDA and private forecasts is the highest for wheat and the lowest for soybeans. |

| **Researcher & Topic (Market Reactions to the USDA forecasts)** | **Summary of study** |
|---|---|
| Aulerich et al. (2007) The Impact of Measurement Error on Estimates of the Price Reaction to USDA Crop Reports. | Implication of Identification by Censoring (ITC) method shows that market reactions to unanticipated information in the USDA forecasts are significantly high. |
| Colling & Irwin (1990) The reaction of live Hogs futures prices to USDA Hogs and Pigs reports. | Live Hogs future prices do not react to anticipated changes in the USDA forecasts, but considerably react to unanticipated changes in the reports. However, the Hogs prices adjust to unanticipated reports on the day following release of the forecasts. |
| Colling et al. (1992) Weak-and strong-form rationality tests of market analysts' expectations of USDA Hogs and Pigs reports. | Expectations of Pigs and Hogs reports are strong-form rational. |
| Colling et al. (1996) Reaction of Wheat, Corn, and Soybean Futures Prices to USDA" Export Inspections" Reports. | Soybean prices respond substantially to unanticipated information in "Export Inspections" reports. Also, corn prices react notably to unanticipated information during the December to February quarter, but soybean prices respond to such an unanticipated information during June to August quarter. |
| Colling et al. (1997)<br>Future price responses to USDA's Cold Storage report. | Live Hogs and pork belly prices react significantly to unanticipated information from the USDA forecasts. Therefore, the forecasts provide information to the markets. |
| Darby (2015)<br>Information Content of USDA Rice Reports and Price Reactions of Rice Futures. | The USDA forecasts provide useful information to the rice markets and rice futures react to the USDA information consistently. |
| Fortenbery et al. (1993)<br>The effects of USDA reports in futures and options markets. | The effects of the USDA forecasts are minimal, but regression tests show that market participants cannot forecast market future. |
| Good & Irwin (2005)<br>Understanding USDA corn and soybean production forecasts: Methods, performance and market impacts over 1970-2005. | The USDA corn and soybeans production forecasts are reasonably well. |
| Irwin at al. (2001)<br>The value of USDA outlook information: an investigation using event study analysis. | The USDA forecasts have significant impacts in soybeans and corn markets. Also, the reports reduce uncertainty of the expected distribution of the prices which improves the market participants' welfare. |
| Isengildina-Massa et al. (2004)<br>Does the Market Anticipate Smoothing in USDA Crop Production Forecasts? | Except for some cases market participants are aware of USDA smoothing practices and efficiently apply this information into their own forecasts. |
| Fortenbery & Sumner (1993)<br>The effects of USDA reports in futures and options markets. | During the time, market participants have learned how to digest the USDA reports. Hence, forecasts do not cause abnormally large price changes. |
| Hoffman et al. (2015)<br>Forecast performance of WASDE price projections for US corn | The USDA WASDE projections of corn season-average price provide valuable information to the market and improves the efficiency of the United States agricultural sector. |
| Karali (2012)<br>Do USDA Announcements Affect Comovements Across Commodity Futures Returns? | On the release days of the grain stocks, feed outlooks, and Hogs and Pigs report the largest movements in covariances happen. |
| McKenzie (2008)<br>Pre-harvest price expectations for corn: The information content of USDA reports and new crop futures. | Results indicate that the USDA forecasts are newsworthy. Also, price reactions to the reports are rational. |
| Patterson & Brorsen (1993)<br>USDA Export Sales Report: Is It News? | The USDA forecast doesn't provide new information to the market and indeed the traders predict the reports. |
| Pruitt et al. (2014)<br>End user preferences for USDA market information. | Results show preference for farm level forecasts by Extension agents. |
| Roberts (2006)<br>The value of plant disease early-warning systems: A case study of USDA's soybean rust coordinated framework | The USDA forecasts provide valuable information to the market. Probably in 2005 the value of information by the USDA forecasts exceeds the cost of getting information. |
| Schroeder et al. (1990)<br>Abnormal returns in livestock | The USDA forecasts do not have consistent upward or downward influences on the prices, but the volatility of returns increases around |
| futures prices around USDA inventory report releases. | the report release time which suggests forecasts provide new information to the market. Also, comparing to the other markets the forecast contains less information for the Hogs market. Hence, the Hogs prices are more volatile after the release of the USDA forecasts. |
| Summer & Mueller (1989) Are harvest forecasts news? USDA announcements and futures market reactions. | There are significant differences between the means and variances of prices following a USDA announcement and the means and variances of prices of the other days. |

**Accuracy of the USDA Forecasts**

As the summery of the relevant studies above show, not all the researchers are on a same page about accuracy of the USDA forecasts. On the one hand some studies maintain that USDA estimates are *unbiased* (e.g. No & Salassi[4] (2009), Sanders & Manfredo[5] (2002), Xiao et al[6]. (2014), Irwin et al[7]. (2014)) and on the other hand, some studies claim that USDA forecasts are *biased* (e.g. Sanders & Manfredo[8] (2003a), Sanders & Manfredo (2003b), Schaefer & Myers (1999)).

Some studies, however, maintain that the USDA forecasts are *inefficient* (e.g. Schaefer & Myers (1999), Sanders & Manfredo (2002). Xiao et al. (2014)), *not optimal* (e.g. Von Bailey & Brorsen (1998), Sanders & Manfredo (2003a), Sanders & Manfredo (2003b)), or *not rational in the long-run* (e.g. Also, Sanders & Manfredo (2008), No & Salassi (2009)).

Some of the studies report an *improvement in accuracy* of USDA forecasts (e.g. Gunnelson et al[9]. (1972), Egelkraut et al. (2003), Good & Irwin[10] (2005), Irwin et al[11]. (2014).

[4] Salassi (2009) argues that USDA forecasts are unbiased in the short-run, but not rational in the long run.
[5] Sanders & Manfredo (2002) maintain that USDA forecasts are unbiased but not efficient.
[6] Xiao et al. (2014) argue that USDA forecasts are unbiased but inefficient.
[7] Irwin et al. (2014) maintain that USDA NASS forecasts for corn are unbiased.
[8] Sanders & Manfredo (2003a) and Sanders & Manfredo (2003b) indicate that USDA forecasts of Broiler price is biased.
[9] Gunnelson et al. (1972) report a moderate improvement in USDA forecasts.
[10] Irwin (2005) report an improvement in accuracy of USDA forecasts for soybeans.
[11] Irwin et al. (2014) maintain that USDA NASS forecasts for corn are improved.

Some of the studies *compare* the accuracy of the USDA forecasts with that of other forecasts (e.g. Kastens et al. (1998), Manfredo & Sanders (2004)). Furthermore, at least two studies indicate that USDA forecasts are *more accurate in case of corn production*, but this is not the case for soybeans production (e.g. Egelkraut et al. (2003), Irwin et al. (2014)).

Fig. 3, part A represents the summary of major findings of the studies that focus on evaluation of accuracy of USDA forecasts. Overall the authors of 4 studies believe that at least for some of the Agriculture products the forecasts are unbiased, 4 studies point out that the accuracy of the forecasts have improved, and 2 studies maintain that USDA does a better job about corn forecasts comparing to soybeans forecasts. However, 3 studies indicate that the USDA forecasts are biased, 3 of them report inefficiency, another 3 studies specify that the forecasts are not optimal, and 2 of them argue that they are not rational.

**Market Reactions to the USDA Forecasts**

Market reactions to the USDA forecasts are not unambiguously identified. While on the one hand some researchers argue that the forecasts are *newsworthy* and provide new and useful information to the market (e.g. Summer & Mueller (1989), Schroeder et al. (1990), Fortenbery et al. (1993), Roberts (2006), McKenzie (2008), Darby (2015), Hoffman et al. (2015)), on the other hand other researchers maintain that the USDA forecast are *not newsworthy* and in fact market participants predict the reports (e.g. Patterson & Brorsen (1993), Isengildina-Massa et al. (2004)).

Also, a couple of studies note that the USDA forecasts *cause market reaction or movement in the prices* (e.g. Colling & Irwin (1990), Colling et al. (1996), Colling et al. (1997), Irwin at al. (2001) (corn and soybeans), Aulerich et al. (2007), McKenzie (2008), Karali (2012)). Furthermore, Colling & Irwin (1990), Colling et al. (1996), Colling et al. (1997), Aulerich et al. (2007) argue that *market reacts to the unanticipated changes* in the forecasts. Fortenbery & Sumner (1993) believe that USDA forecasts *do not cause uncertainty*. In addition, Colling et al. (1992) maintain that expectations of Pigs and Hogs reports are *strong-form rational*. Some other researchers such as McKenzie (2008) claim that *reactions to prices are rational*.

Fig. 2, part B represents the summary of major findings of the studies that focus on the market reactions to the USDA forecasts. All in all, 2 studies claim that the forecasts are not newsworthy, while 7 of them argue that they are newsworthy. 7 studies specify that USDA forecasts cause market reactions, 4 of them maintain that markets react to unanticipated information, 2 studies argue that market expectations are rational, and 1 study maintain that the forecasts don't cause uncertainty.

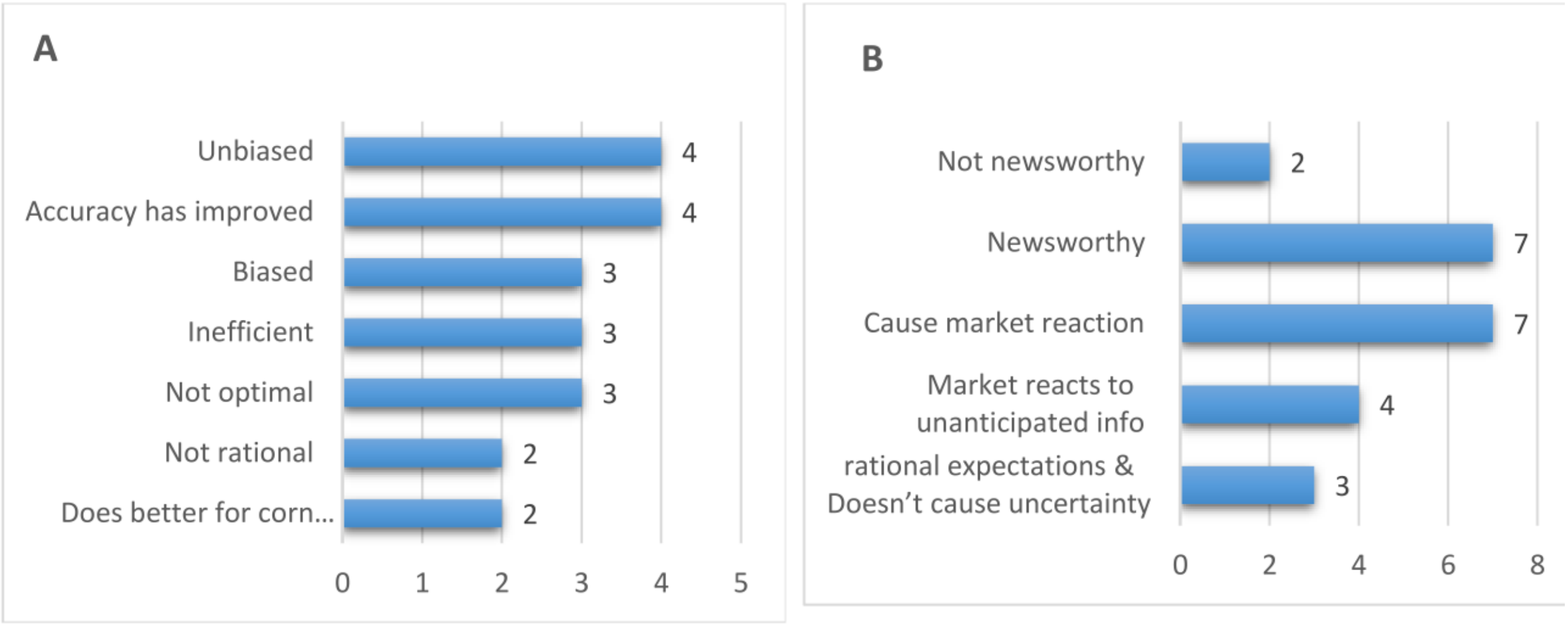

**Fig. 2. Summary of the major findings of the published studies.** A represents a summary of main findings of the studies which focus on analyzing the accuracy of the USDA forecasts, while B shows the ones which study market reactions to the USDA forecasts.

## Meta-analysis

A possible problem with the USDA forecasts can be repeating the past errors or over-correcting them. A correlation with the past errors represents the forecasts tendency to repeat or overcorrect the past errors. Positive correlation with past forecasts means that the new forecasts repeat the same errors, while negative correlation represents over-correction of the errors (Isengildina-Massa et al, 2013). Some of the studies calculate the Pearson correlation of the USDA forecasts with their past errors (e.g. Sanders & Manfredo, 2002 and 2003; Isengildina-Massa et al., 2004, 2006, 2012, and 2013; Good & Irwin, 2005; and McKenzie, 2008). We apply their findings which are represented in Table. 1. to do the meta-analysis in this study. Note that AR4 which is a time series model represents a substitute method of forecasting suggested by the authors.

In a meta-analysis study usually two models get discussed: fixed-effect and random-effect models. In a fixed-effect model the assumption is that the dataset in not random and the individuals are from a same population while in random effect models the dataset is from a hierarchy of different populations and the differences among the dataset observations relates to that hierarchy. As an example, the dataset which is collected from a same population in a same library may qualify for the fixed-effect model. The fixed-effect model doesn’t account for heterogeneity and if indeed the dataset is from different populations it overestimates the effect sizes. In that condition, someone may apply the random-effect model. When there is heterogeneity in the dataset the calculated Confidence Intervals (CI) are much wider if the researcher applies the random-effect models, but if the dataset is homogeneous the CI is same as the estimated CI using fixed-effect models.

| Table. 1. The Dataset to do meta-analysis | | | | | | |
|---|---|---|---|---|---|---|
| | Authors | Year | Time Period | Item | Pearson Correlation | Forecast |
| 1 | Sanders & Manfredo | 2002 | 1982-2000 | beef | 0.31 | USDA |
| 2 | Sanders & Manfredo | 2002 | 1982-2000 | pork | 0.15 | USDA |
| 3 | Sanders & Manfredo | 2002 | 1982-2000 | broiler | 0.25 | USDA |
| 4 | Sanders & Manfredo | 2002 | 1982-2000 | beef | -0.12 | AR4 |
| 5 | Sanders & Manfredo | 2002 | 1982-2000 | pork | -0.02 | AR4 |
| 6 | Sanders & Manfredo | 2002 | 1982-2000 | broiler | 0.03 | AR4 |
| 7 | Sanders & Manfredo | 2003 | 1982-2002 | cattle | 0.24 | USDA |
| 8 | Sanders & Manfredo | 2003 | 1982-2002 | Hogs | 0.18 | USDA |
| 9 | Sanders & Manfredo | 2003 | 1982-2002 | broiler | 0.31 | USDA |
| 10 | Sanders & Manfredo | 2003 | 1982-2002 | cattle | 0.02 | AR4 |
| 11 | Sanders & Manfredo | 2003 | 1982-2002 | Hogs | -0.21 | AR4 |
| 12 | Sanders & Manfredo | 2003 | 1982-2002 | broiler | 0.17 | AR4 |
| 13 | Isengildina et al. | 2004 | 1970-2002 | corn | 0.45 | USDA |
| 14 | Isengildina et al. | 2004 | 1970-2002 | soybeans | 0.22 | USDA |
| 15 | Good & Irwin | 2005 | 1970-2005 | corn | 0.54 | USDA |
| 16 | Good & Irwin | 2005 | 1970-2005 | soybeans | 0.35 | USDA |
| 17 | Isengildina et al. | 2006 | 1970-2002 | corn | 0.23 | USDA |
| 18 | Isengildina et al. | 2006 | 1970-2002 | soybeans | -0.8 | USDA |
| 19 | McKenzie | 2008 | 1970-2005 | corn | 0.66 | USDA |
| 20 | Isengildina et al. | 2012 | 1985-2009 | corn | -0.31 | USDA |
| 21 | Isengildina et al. | 2013 | 1987-2010 | soybeans | 0.11 | USDA |
| 22 | Isengildina et al. | 2013 | 1987-2010 | wheat | 0.16 | USDA |

To determine heterogeneity in the sample sizes we calculate Q-statistic. The null hypothesis for the Q-statistic test is that ‘all of the studies share a same effect size’ and the alternative hypothesis is that ‘the studies do not examine a common effect size’. In other words, a statistically Q-statistic means that the studies do not share a common effect size. However, a non-significant Q-statistic doesn’t prove that the dataset is homogeneous. The test for heterogeneity results show that Q-statistic is 77.3 and p-value < 0.0001 which means that the studies do not share a common effect size and the dataset is heterogeneous.

An alternative test for heterogeneity applies I2-statistic. I2-statistic is a percentage that shows that the proportion of variance is from actual differences between studies rather than within the study variance. Higgins et al. (2003) provide thresholds of 25%, 50%, and 75% which indicate low, moderate and high variance for I2-statistic. For our dataset I2-statistic is 70.3% (95% CI: 48.5, 83.8) which represents moderate to high variance.

Even though, the mentioned tests show that there is heterogeneity in the dataset, but they don't provide any clue that which studies may disproportionally affect heterogeneity. Instead, Baujat plot which introduced by Baujat et al. (2002) makes it possible to see which studies contribute to the heterogeneity and overall influence the results more than the others. For the mentioned plot the horizontal axis shows the study heterogeneity while the vertical axis indicates the influence of studies on the overall results. Fig. 3. represents Baujat plot for our dataset.

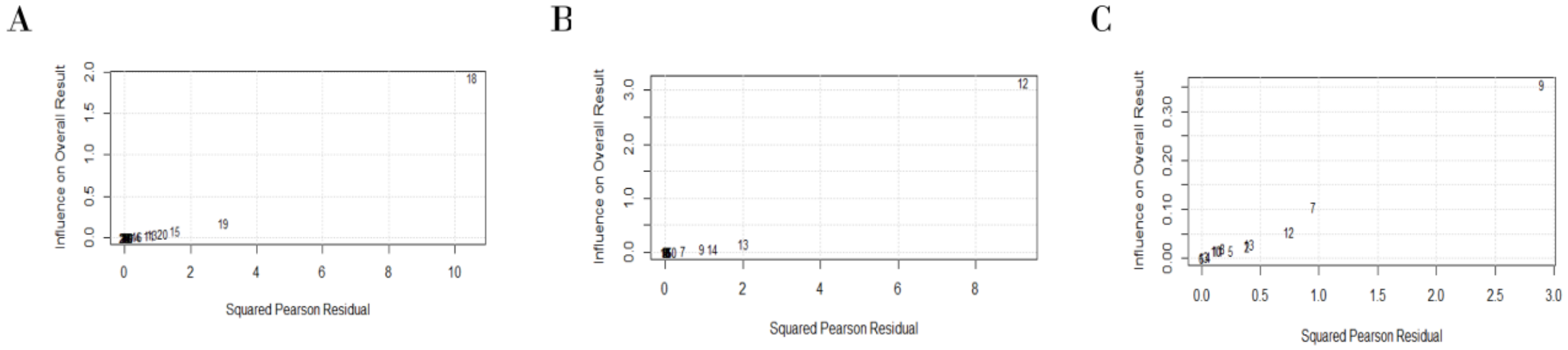

**Fig. 3. Baujat plot to identify the studies that contribute to heterogeneity**. Each number represents a study. Studies on top right have greater influence on the results and have a bigger contribution to heterogeneity. plot A considers all of the studies. As can be seen in the graph, study 18 which is Isengildina-Messa et al. (2006) for soybeans contributes the most to heterogeneity. In plot B, the AR4 models are eliminated and only the studies which focus on USDA forecasts are left. Here study 12 is in the right corner above. In plot C the studies with biggest variation and small size effects are eliminated.

Another important concept in meta-analysis literature is publication bias which represents that the studies with stronger effect-sizes are more probable to get published. In other words, the publisher looks at the findings of the research and the studies with strong and positive results have more chances to get published. Funnel plot is a helpful tool to determine publication bias. In this plot the vertical axis shows individual effect sizes while the horizontal axis represents standard errors. A symmetric Funnel plot indicates the possibility of unbias publication while an asymmetric plot shows the possibility of publication bias. If the plot shows a negative correlation that means probably the studies with small and negative results do not get published and they miss from the left corner of the plot. Fig. 4. represents Funnel plot for our dataset. As can be seen in most of the cases the plot shows positive correlations.

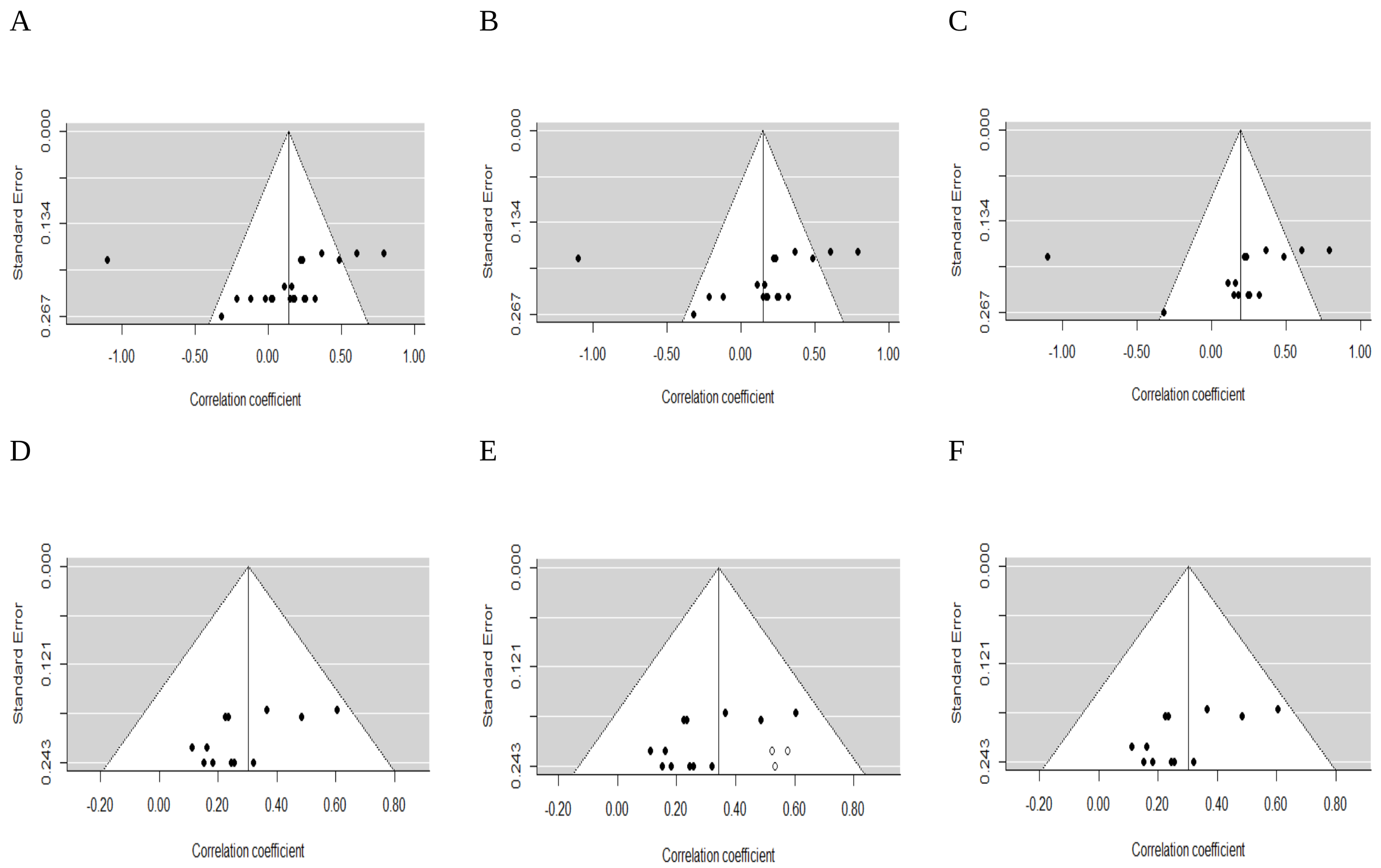

**Fig. 4. Funnel Plot to represent publication bias.** plot A which includes all of the studies in Table. 1. shows a positive correlation and therefore the dataset can be interpreted as asymmetric. In plots B and C, we remove the studies with small effect sizes and big variations. Funnel Plot D includes all of the studies in plot A except the AR4 models. Plot E simulates three removed studies of plot D which if they were there the plot would be symmetric. In Funnel Plot F, the studies with small effect sizes and big variations are removed from Plot D which again sounds like an asymmetric plot. Overall, it sounds that Funnel plot in all of the scenarios is asymmetric which demonstrates the possibility of publication bias.

A weakness of Funnel plot is that it is only a subjective measure of possibility of publication bias. We apply Rank Correlation and Egger's tests as objective tools to test for publication bias. Begg and Mazumdar (1994) propose Rank Correlation test. Based on their method $P<0.05$ is consistent with asymmetrical Funnel plot. However, Rank Correlation test cannot be fully trusted for studies with less than 25 studies (Sterne at al. 2000). An alternative test which is more useful for meta-analysis with less than 25 studies is Egger's test represented by Egger et al. (1997). Here, Rank Correlation test (p = 0.0081) is statistically significant which suggests that there is publication bias in our dataset. However, based on the Egger's test which is another tool to test for funnel plot asymmetry, p equals to 0.2408 which is not statistically significant. Unlike Rank Correlation test, this finding suggests that the studies are not symmetric in the Funnel plot. In other words, based on the results of Egger's test there in no evidence of publication bias.

**Discussion**

By combining the findings of a variety of studies, providing useful statistical tests and procedures, and aggregating and synthesizing the results Meta-analysis helps to get resolve uncertainty when the studies contradict and it is certainly useful to get a vivid and pig picture of findings of many studies in one place.

Many researchers have studied USDA forecasts, but almost all of the academic publications in this area can be divided in two groups. The studies which evaluate the accuracy of the USDA forecasts and the ones that evaluate the market reactions to the USDA forecasts. Some of the studies do both. These groups of studies provide a variety of results and in many cases their findings contradict. Therefore, in this study we do a meta-analysis on the published studies in this area to tackle the following questions:

1) how the academic published studies evaluate accuracy of the USDA forecasts?

2) how the academic published studies evaluate market reactions to the USDA forecasts?

3) Is there heterogeneity in the results of the studies?

4) Is there any publication bias in the published studies in this area?

After aggregating and synthesizing all published papers that we could find, we figured out that some of the studies maintain that the forecasts are unbiased, while most of the studies point out that at least for some of the products the USDA forecasts are not efficient, they are biased, and they are not optimal.

About market reactions to the USDA forecasts, we found a few studies that claim that the forecasts are not newsworthy, and the market participants could predict the reports. However, most of the studies argue that the forecasts are newsworthy, they provide useful information to the market participants, and they cause market reactions and change in the prices. We did meta-analysis using a package named “metaphor” in R to tackle the third and the fourth questions. We applied Q-statistic, I2-statistic, and Baujat plot to test for heterogeneity. Based on the findings from the mentioned tests the results of the studies are heterogeneous. Also, we applied Funnel plot, Rank Correlation test, and Egger’s test to test for publication bias. Funnel plot and Rank Correlation test results show publication bias. However, as we already mentioned Egger’s test findings are more accurate for small datasets and the results of this test doesn’t confirm publication bias.